\documentclass[graybox]{svmult}

\usepackage{mathptmx}       % selects Times Roman as basic font
\usepackage{helvet}         % selects Helvetica as sans-serif font
\usepackage{courier}        % selects Courier as typewriter font
\usepackage{makeidx}         % allows index generation
\usepackage{graphicx}        % standard LaTeX graphics tool
\usepackage{multicol}        % used for the two-column index
\usepackage[bottom]{footmisc}% places footnotes at page bottom
\usepackage{subfigure}
\usepackage{indentfirst}

\usepackage[misc]{ifsym} %used for the envelope
\makeindex             % used for the subject index
\begin{document}

\title*{Investigating the effect of social groups in uni-directional pedestrian flow}
\author{Luca Crociani, Yiping Zeng, Andrea Gorrini, Giuseppe Vizzari, Weiguo Song}
\authorrunning{Crociani, L., Zeng, Y., Gorrini, A., Vizzari, G., Song, W.}
\institute{Luca Crociani \and Andrea Gorrini \and Giuseppe Vizzari \at CSAI research center, University of Milano - Bicocca, Milano, Italy\\\email{\{name.surname\}@disco.unimib.it}
\and Yiping Zeng (\Letter) \and Weiguo Song \at Sate Key Laboratory of Fire Science, University of Science and Technology of China, Hefei, China\\\email{ypzeng@mail.ustc.edu.cn}
\and Weiguo Song\\\email{wgsong@ustc.edu.cn}}

%\author{Luca Crociani, Andrea Gorrini, Claudio Feliciani, Giuseppe Vizzari, %Katsuhiro Nishinari, Stefania Bandini}
%\authorrunning{Crociani, L., Gorrini, A., Feliciani, C., Vizzari, G., Nishinari, K., Bandini, S.}
%\institute{Luca Crociani(\Letter) \and Andrea Gorrini \and Giuseppe Vizzari \and Stefania Bandini \at CSAI research center, University of Milano-Bicocca, Milan, Italy.\\\email{\{name.surname\}@disco.unimib.it}
%\and Andrea Gorrini 
%\\\email{andrea.gorrini@disco.unimib.it}
%\\ Giuseppe Vizzari
%\\\email{giuseppe.vizzari@disco.unimib.it}
%\\ Stefania Bandini \at\email{stefania.bandini@disco.unimib.it} 
%\and Claudio Feliciani \and Katsuhiro Nishinari \and Stefania Bandini \at RCAST, The University of Tokyo, Tokyo, Japan.
%\\\email{claudio.feliciani@gmail.com}
%\and Katsuhiro Nishinari
%\\\email{tknishi@mail.ecc.u-tokyo.ac.jp}
%}

\maketitle

\abstract{The influence of cohesion among members of dyads is investigated in scenarios characterized by uni-directional flow by means of a discrete model: a corridor and the egress from a room with a bottleneck of varying width are simulated. The model manages the dynamics of simulated group members with an adaptive mechanism, balancing the probability of movement according to the dispersion of the group; the cohesion mechanism is calibrated through the parameters $\kappa_c$ and $\delta$. All scenarios are simulated with two procedures: (Proc. 1) population composed of individual pedestrians, in order to validate the simulation model and to provide baseline data; (Proc. 2) population including dyads (50\% of the simulated pedestrians), in order to verify their impact. In the corridor scenario, the presence of dyads causes a reduction of the velocities and specific flow at medium-high densities. Egress from a square room with a unique central exit produces results in line with recent studies in the literature, but also shows that the dyads negatively affect the dynamics, leading generally to a slower walking speed and a lower pedestrian flow. Ignoring the presence of dyads would lead to an overestimation of egress flows.}

\setlength{\parindent}{2em}
\section{Introduction}
\label{sec:1}
Recently, the topic of social groups has gathered a growing interest from researchers studying the dynamics of pedestrians. One one side, field observations (e.g.~\cite{gorrini2016age}) have highlighted that a crowd of pedestrians is mostly composed of groups of up to 4 members (dyads are typically the most frequent) and that people walking in a group are generally slower than individuals. Controlled experiments involving groups have been designed to investigate their aggregated effect, and a relevant impact is observed in a corridor setting with bi-directional flow (see, e.g.~\cite{gorrini2016social,crocianiTGFcouplesExp}) and a bottleneck scenario(see, e.g.~\cite{von2017empirical}). On the modelling side, many works are being proposed for the simulation of group behaviour: for example, the work in~\cite{PhysRevE.89.012811} proposes a continuous model for the behavior of groups of 2 or 3 members based on field observations, while~\cite{muller2016evacuation} design leader-follower structure of dyads using dynamic floor fields. In this paper, we present a discrete model considering groups, based on the work in~\cite{DBLP:journals/jca/BandiniCV17}, which is extended and calibrated to fit the data from a controlled experiment~\cite{gorrini2016social}. The calibrated model is then used to investigated the aggregated effect of dyads in uni-directional pedestrian flows at high densities (Sec.~\ref{sec:FD}) and passing through a bottleneck (Sec.~\ref{sec:bottleneck}).

%Lu et al. \cite{lu2017study} divided groups into follows and leaders
%In the research field of social relationship, most groups are composed of two (dyads) or three occupants \cite{zanlungo2015spatial,zanlungo2017effect}. So the majority of scholars investigate the social group with two or three people in respect of experiments and modellings. Experiments could provide lots of basic characteristics of occupants and modellings could help researchers reduce the cost of investigations and provide relative results. Andrea et al. \cite{gorrini2016social} carried out counterflow experiments and found that dyads walked significantly slower than singles, due to the difficulty in movement coordination among group members. As for the modelling, However, fewer attention is paid to the cohesion between dyads and the effect of dyads and this cohesion has a relationship with velocity and behaviours of occupants. In this paper, this work presents an extension of a model considering cohesion and groups at the operational level. And we focus on the movement of dyads in a corridor and the influence of dyads on the outflow with a bottleneck.

\section{A Model for Group Cohesion}
\label{sec:2}
The simulation model here presented is designed to achieve a more realistic simulation of pedestrian and group dynamics, with particular attention to the shape of the group. The model is discrete in space and time, and at each time-step of the simulation agents evaluate cells $c$ of the Moore neighbourhood with the utility function $U(c)$. This aggregates the components associated to the reproduction of a particular behaviour by means of a weighted sum:

\begin{footnotesize}
\begin{equation}
U(c)=\frac{\kappa_g G(c)+\kappa_{ob} Ob(c)+\kappa_s S(c)+\kappa_{c} C(c)+\kappa_{d} D(c)+\kappa_{ov} Ov(c)}{d}
\end{equation}
\end{footnotesize}

Individual functions model respectively: (i) goal attraction; (ii) obstacle repulsion; (iii) keeping distance from other pedestrians; (iv) cohesion with other group members; (v) direction inertia; (vi) overlapping to avoid gridlock in counter-flow situations. The first three elements are modelled with the usage of the well-known \emph{floor-field} approach to model the base behaviour of pedestrians: movement towards a target, obstacle avoidance, proxemics with other pedestrians in a repulsive sense.

The function $C(c)$ is introduced to manage the cohesion among group members and its strength is mainly adjustable by the parameter $\kappa_c$. The function calculates the level of attractiveness of each neighbour cell, according to the position and velocity of the other agents of the group. These information are used to estimate their next positions and then evaluate the attractiveness of each cell according to the cohesion, as the following:

\begin{equation}\label{eq:c}
C(c) = \left[ \left( \eta \cdot \sum_{\hat{a} \in G\setminus \{a\}}\frac{\|\vec{x}_{a}-\vec{x}_{\hat{a}}^\bullet\| - \|\vec{x}_{c} - \vec{x}_{\hat{a}}^\bullet\|}{|G|-1}\right) \cdot 2 \right] - 1
\end{equation}

\begin{figure}[t]
\begin{center}
\subfigure[]{\includegraphics[height=3.5cm]{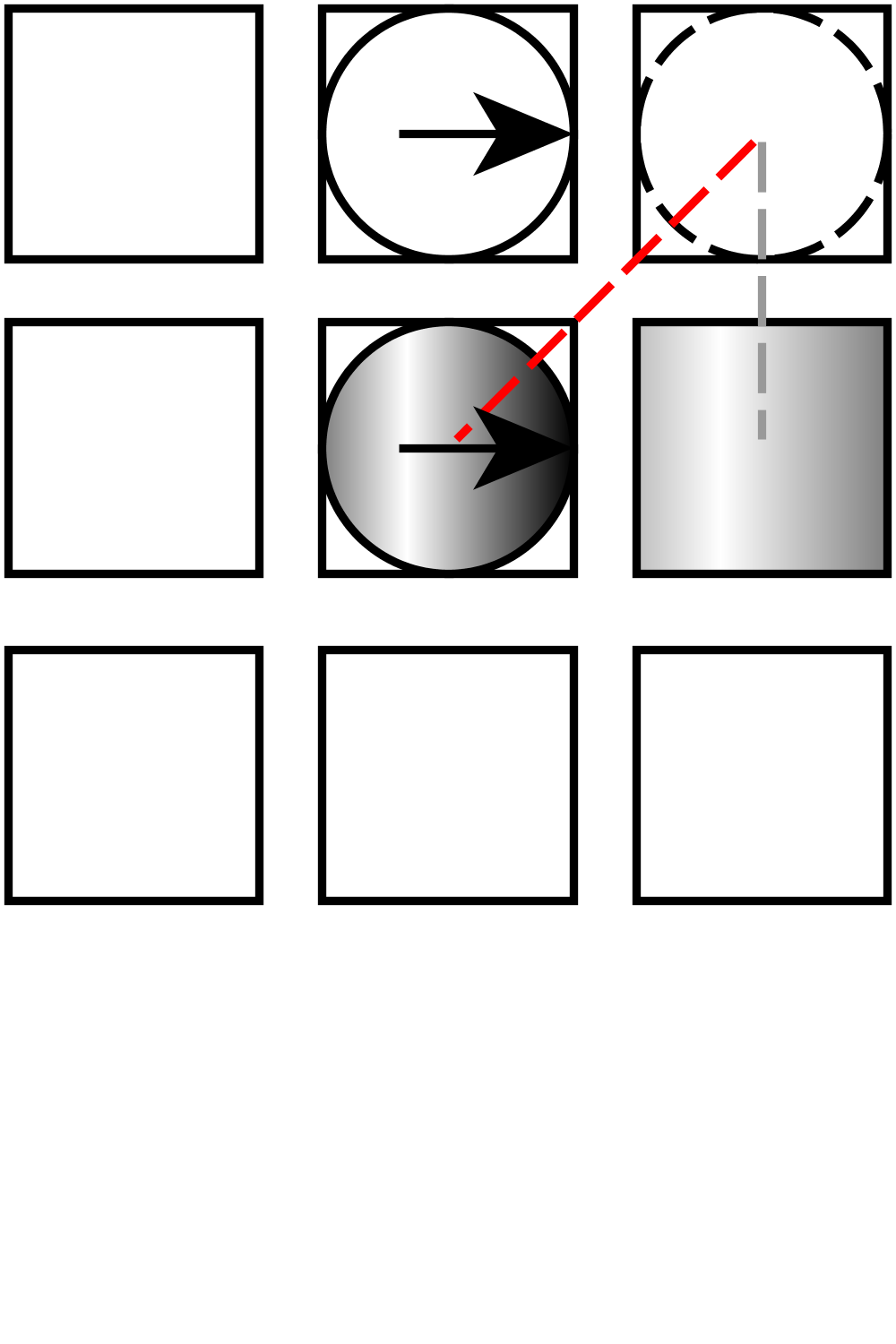}\label{fig:C_c}}
\hspace{1.5cm}
\subfigure[]{\includegraphics[height=4.5cm]{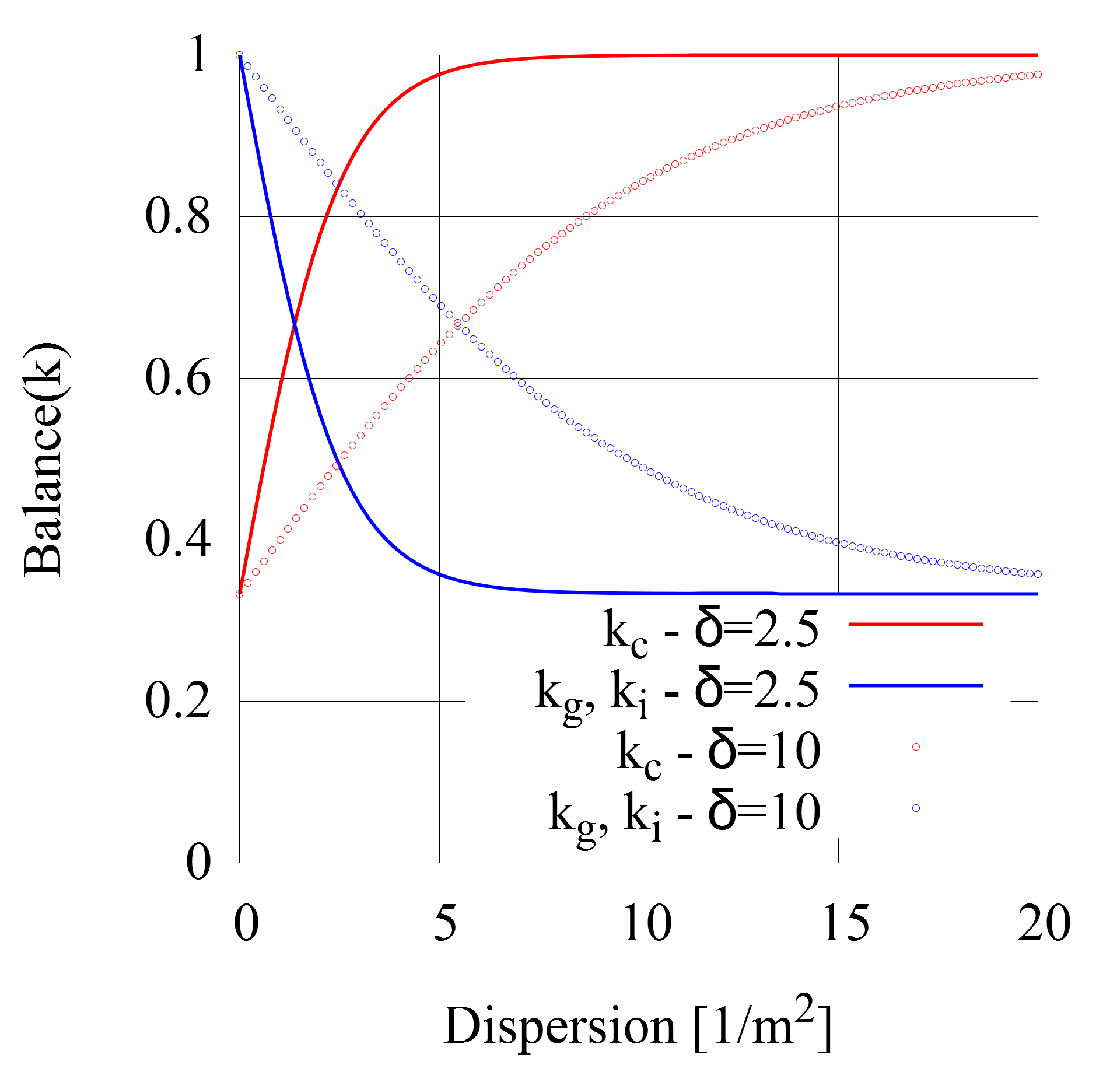}\label{fig:group_balance}}
\caption{(a) Example of calculation of $C(c)$ for a 2-members group. Cells ahead both pedestrians will be the most desirable according to the function. (b) Example of values of the function $Balance$ for the different calibration weights, configuring different values of $\delta$.}
\end{center}
\end{figure}

Where $\eta$ is a normalization factor that, along with numerical values, allows to translate the cohesion value into the range $[-1,1]$. $|G|$ defines the size of the group and $\|\vec{x}\|$ denotes the 2-norm of a vector $\vec{x}$. $\vec{x}_{\hat{a}}^\bullet$ describes a prediction of the next position of the agent $\hat{a}$ and it represents a small advancement to the model that allowed to improve the plausibility of the simulated behaviour. The estimation is calculated using the velocity vector $\vec{v}_{\hat{a}}$: $\vec{x}_{\hat{a}}^\bullet = \vec{x}_{\hat{a}} + \vec{v}_{\hat{a}}$. The functioning of the function is exemplified in Fig.~\ref{fig:C_c}.

By acting on the parameters of $U(c)$ during the simulation it is possible to describe different states of the same pedestrian in different moments of a single simulated scenario. This strategy is applied to allow group members to adapt their behaviour in dense situations or in presence of obstacles. According to this method, weights $\kappa_g, \kappa_{c}, \kappa_{i}$ are varied and possibly inhibited according to an index that describes the instantaneous \textit{dispersion} of the group, as shown in Fig.~\ref{fig:group_balance}. This \emph{adaptation} of the behaviour of group members can be calibrated according to the parameter $\delta$ that, together with $\kappa_c$, will be investigated for the calibration of the model in the next Section. 

After the utility evaluation for all the cells of the neighbourhood, the choice of action is decided by the probability to move in each cell $c$ as ($N$ is the normalization factor): $P(c) = N \cdot e^{U(c)}$.

\section{Calibration of the Group Behaviour}

In order to calibrate the model and reproduce plausible movement of pedestrians and groups, data from the first procedure of a controlled experiment studying the movement of dyads~\cite{gorrini2016social} have been used. In the experimental procedure, a uni-directional pedestrian flow composed of 30 individuals and 24 dyad members\footnote{Dyads were artificially and randomly formed at the beginning of the iteration by asking participants to try to walk close to their companion as they were friends.} crossed a corridor-like setting of 3m width. Trajectories were automatically extracted in a measurement area of 10m length located at the center of the corridor, after a buffer zone of 2m that allowed participants to reach a stable speed. 

\begin{figure}[t]
\begin{center}
\subfigure[]{\includegraphics[width = 0.4\textwidth]{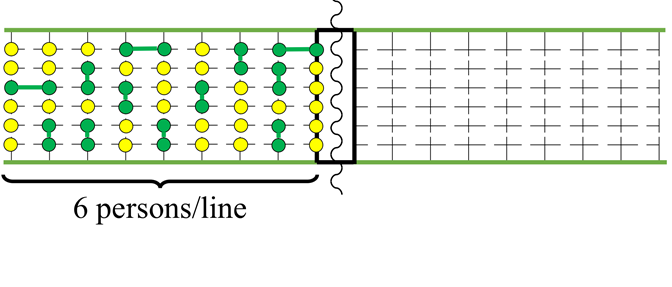}}
\subfigure[]{\includegraphics[width = 0.55\textwidth]{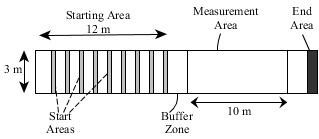}}
\caption{Example configuration of the first procedure of the controlled experiment in~\cite{gorrini2016social} (a) and its realization with the simulation model (b).}
\label{fig:calibration_scenario}
\end{center}
\end{figure}

As shown in Fig.~\ref{fig:calibration_scenario}, a corridor-like setting similar to the experiment has been designed with the simulation model, with an analogous initial configuration of the pedestrian flow: starting positions are given by 9 start areas which generates 6 pedestrians each. Members of dyads are generated close to each other to avoid bias given by the initial configuration. Moreover, only information about pedestrians inside the measurement area are analysed. Given that all participants were young male student, we configured a \emph{desired speed} of simulated pedestrians of 1.6 m/s. 

The aim of the simulation campaign was to calibrate parameters $\delta$ and $\kappa_c$, managing the group behaviour, to fit the results about average speeds of all participants and the distribution of relative positions of dyads, which describes their spatial behaviour during the experimental iteration. To gather stable results, a set of 100 simulation iterations is configured for each configuration of the two parameters that, as pure assumption, is explored in the range $\left[0:30\right]$ with an increase rate of 1. 

\begin{figure}[t]
\begin{center}
\subfigure[]{\includegraphics[width=0.4\textwidth]{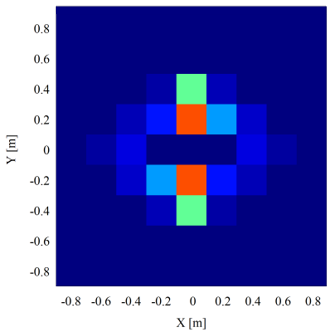}\label{fig:calibration_relPos_obs}}\hspace{.5cm}
\subfigure[]{\includegraphics[width=0.4\textwidth]{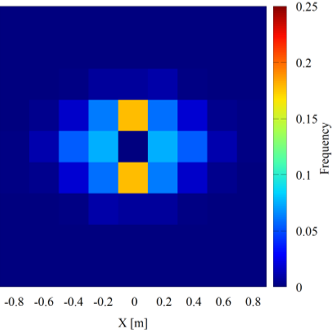}\label{fig:calibration_relPos_sim}}
\caption{Distribution of relative positions of dyads (direction of movement towards right) of the experiment (a) and simulation (b). The same color-scale is applied in both pictures.}
\end{center}
\end{figure}

The calibration phase led to the optimal configuration of parameters $(\delta, \kappa_c) = (7, 12)$, which generates average speeds of singles, dyads and of the overall population as shown in Tab.~\ref{tab:calibration_speeds}. Results are in line with the observation. By looking at the distribution of relative positions (Fig.~\ref{fig:calibration_relPos_obs} and~\ref{fig:calibration_relPos_sim}), the difference is a bit more marked than with the speed, but overall the results are close and they highlights that also in the simulations the most frequent pattern was the line-abreast one.

\begin{table}[t]
\begin{center}
\begin{tabular}{|c|c|c|c|}
 \hline
 & Individual & Dyads & Population \\
 \hline
 \hline
 Experiment & 1.32 & 1.30 & 1.31\\
 \hline
 Simulations & 1.308 & 1.305 & 1.3067\\
 \hline
\end{tabular}
\caption{Comparison of average speeds [m/s] of pedestrians in the experiment and in the simulations.}\label{tab:calibration_speeds}
\end{center}
\end{table}

\section{Validation at basic movement}\label{sec:FD}
\begin{figure}[t]
\begin{center}
\includegraphics[scale=0.3]{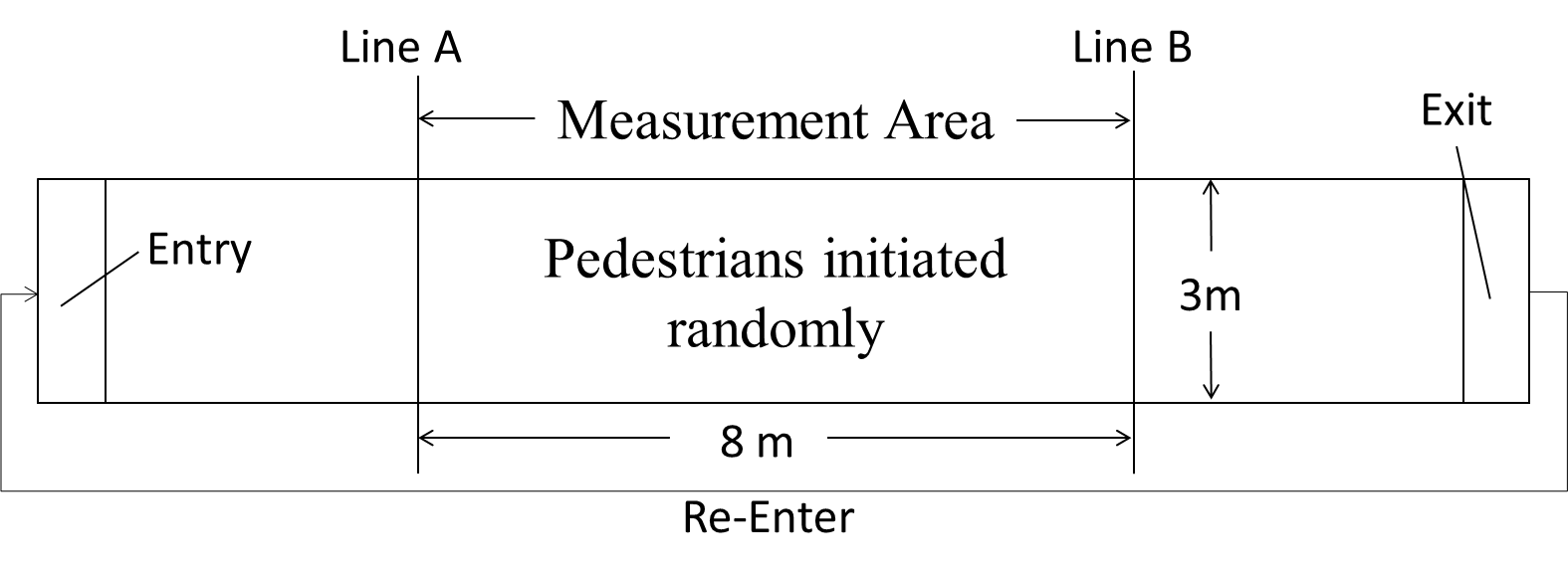}
\caption{The corridor environment used for the test of the fundamental diagram.}
\label{fig:FD-environment}
\end{center}
\end{figure}

In this section, we will study the effects of the presence of dyads on the specific flow in a corridor environment with uni-directional movement. A graphic description of the setting is shown in Fig.~\ref{fig:FD-environment}. Two case-studies have been configured: a baseline with only singles and one with 50\% presence of dyads. To compute the fundamental diagram, simulation campaigns of 10 iterations of 5000 steps each have been run per each investigated average density of pedestrians in the whole environment. The environment was configured as toroidal, so that the number of pedestrians is constant during each iteration. To gain precision with the analysis, data are generated within a \textit{measurement area} of 8 m length, where entrance/exit events are recorded for each agent. This allowed to compute the number and the average speed of pedestrian inside the measurement area for a given time window and to achieve larger datasets with every simulation. Finally, at each run we considered only data generated from step 2000, where a steady state was generally reached.

\begin{figure}[t]
\centering
\subfigure{\includegraphics[width=0.49\textwidth]{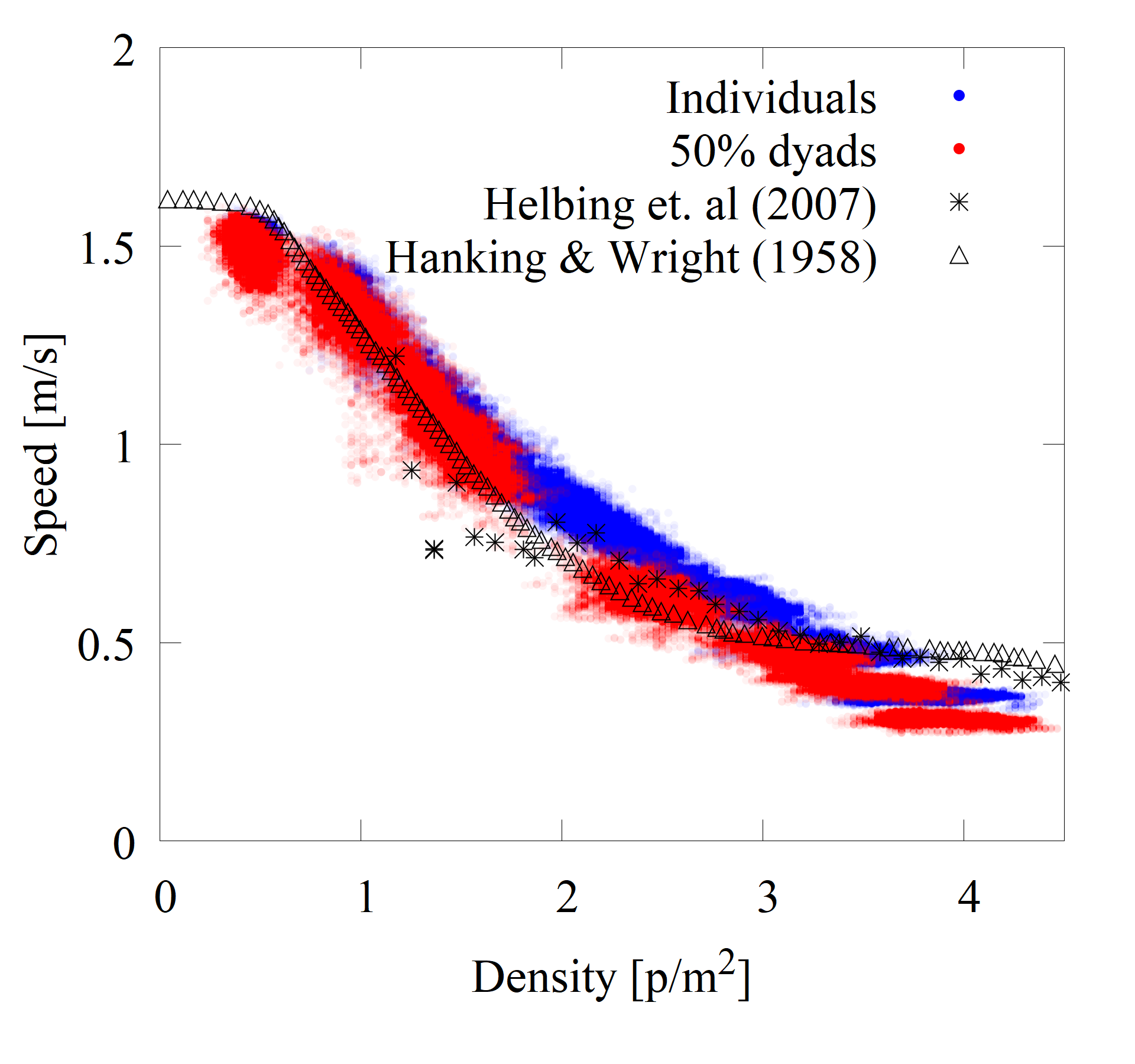}}
\subfigure{\includegraphics[width=0.49\textwidth]{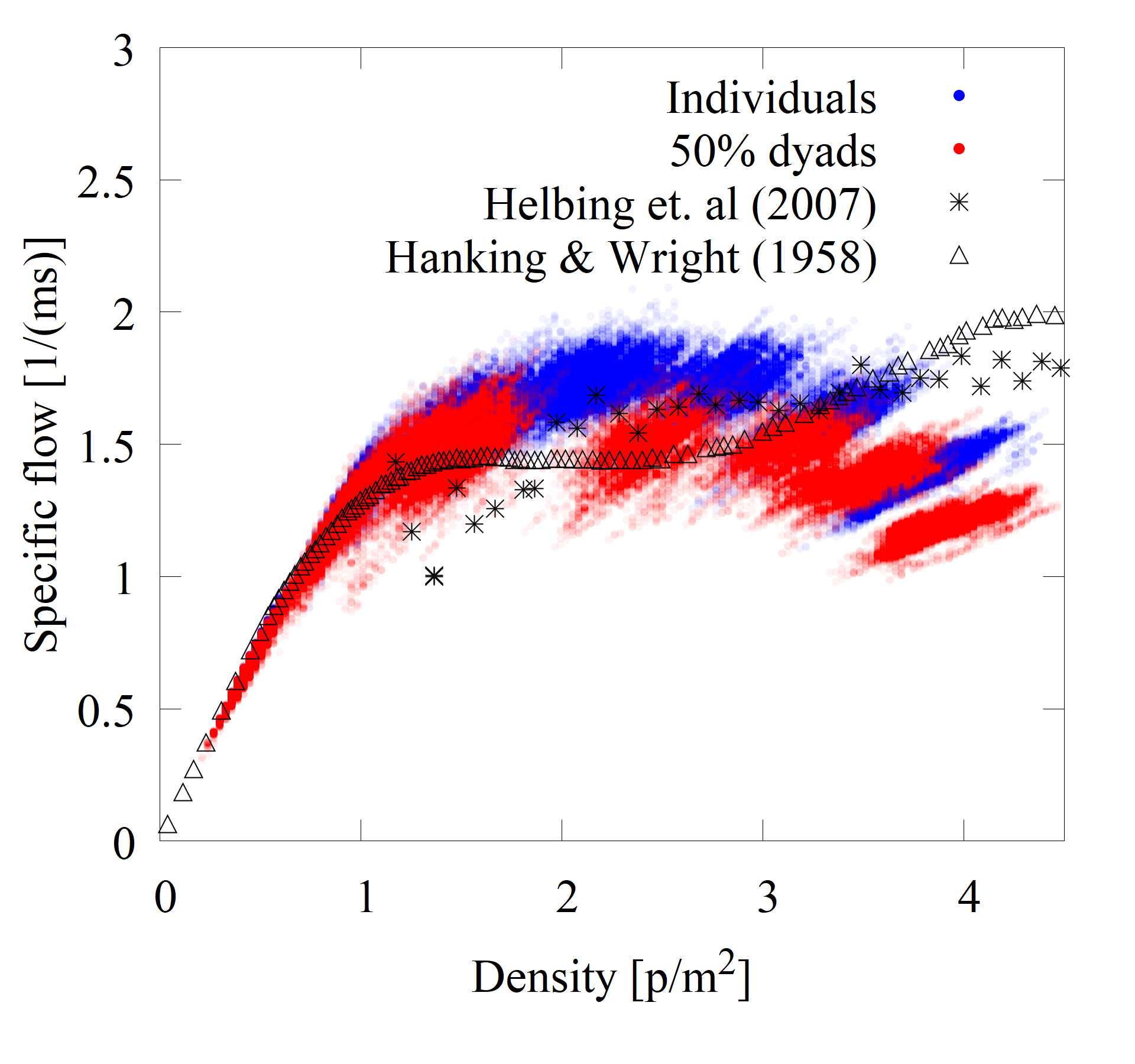}}
\caption{Results of the fundamental diagram in the form density--speed (left) and density--flow (right).}
\label{fig:FD-results}
\end{figure}

Results in the form density--speed and density--flow are shown in Fig.~\ref{fig:FD-results}. Both datasets are in agreement with empirical guidelines and datasets from the literature. At the same time, the specific flow of dyads is smaller than that of individuals for densities higher than 1.5 p/m$^{2}$. On one hand, dyads walk slower than individuals in low density situations, but the difference is quite limited with a similar configuration of their desired speed. The increase of density, on the other hand, has a stronger effects on the average speed of the population with the presence of dyads, emphasizing that dyads had a negative effect on the observed dynamics. This is due to the cohesion mechanism configured for the walking behaviour of dyads, which makes them to keep short distances to each other and to slow down while congested situations leads to a fragmentation of the group. While this behaviour is assumed for this model and it could be considered rather strong compared to the reality, a similar difference among fundamental diagrams observed with only individuals and with about 40\% of dyads was observed in an experiment also presented in this book~\cite{crocianiTGFcouplesExp}.

\section{Analysis of group influence on a bottleneck scenario}\label{sec:bottleneck}
\begin{figure}[t]
\centering
\subfigure[]{\includegraphics[width=0.45\textwidth]{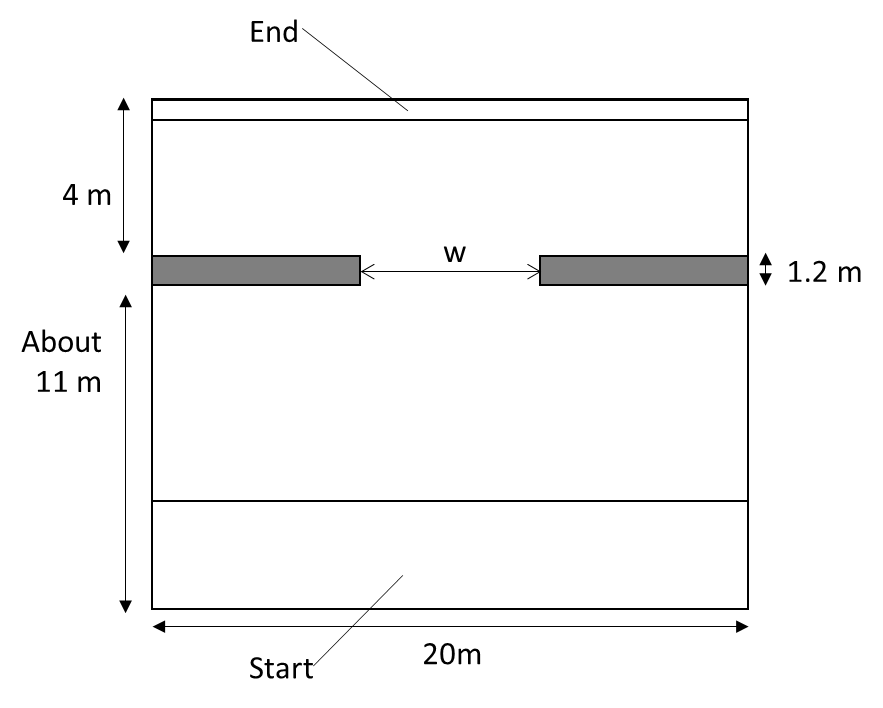}\label{fig:bottleneck_scenario}}
\subfigure[]{\includegraphics[width=0.4\textwidth]{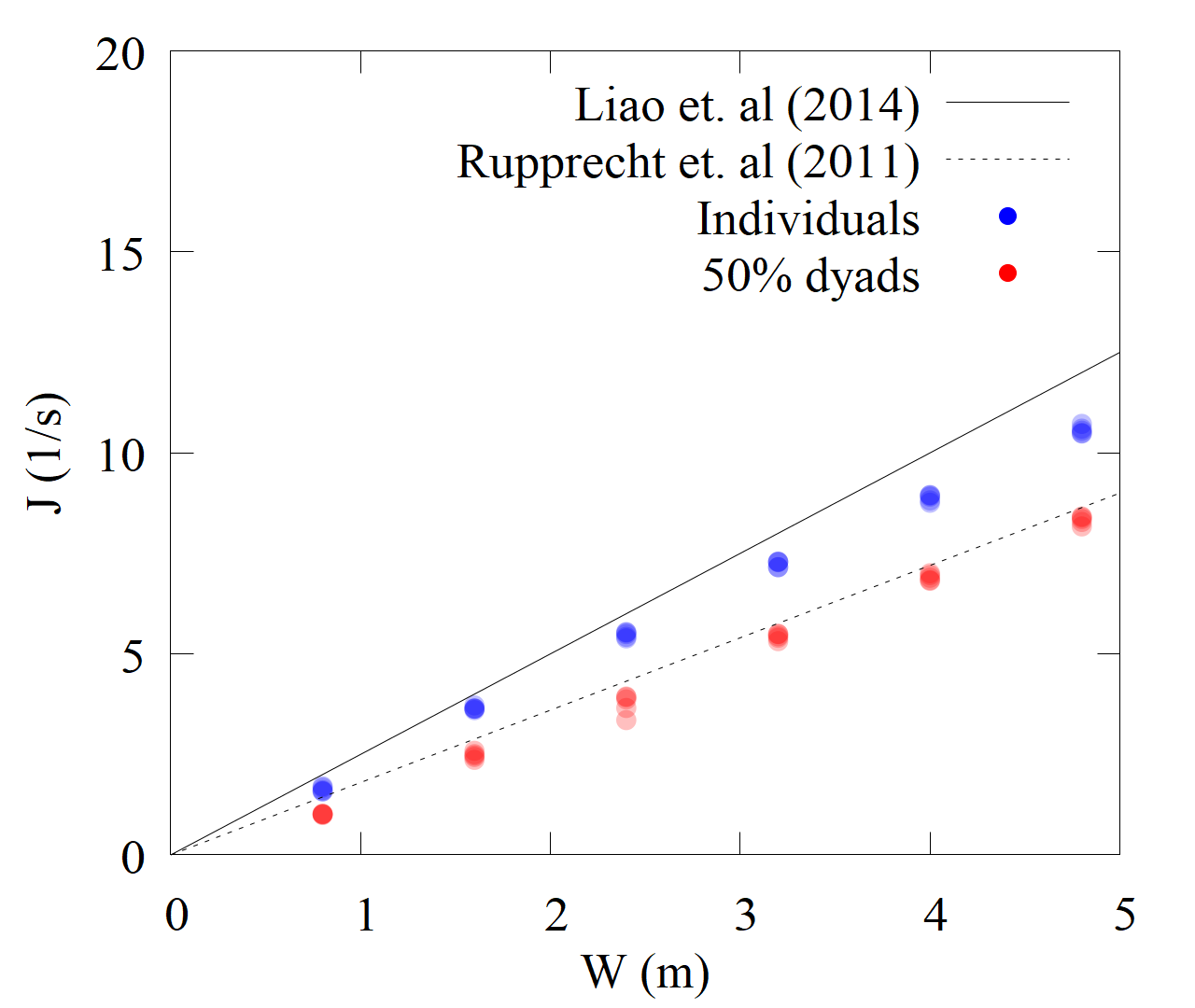}\label{fig:bottleneck_results}}
\caption{(a) Simulated scenario to test the outflow from the bottleneck. (b) Total flow achieved in the two case studies and two empirical datasets from the literature~\cite{liaoPED2014,Rupprecht2011} for comparison.}
\label{fig:implicationscenario}
\end{figure}

\begin{figure}[t]
\centering
\subfigure[]{\includegraphics[width=0.45\textwidth]{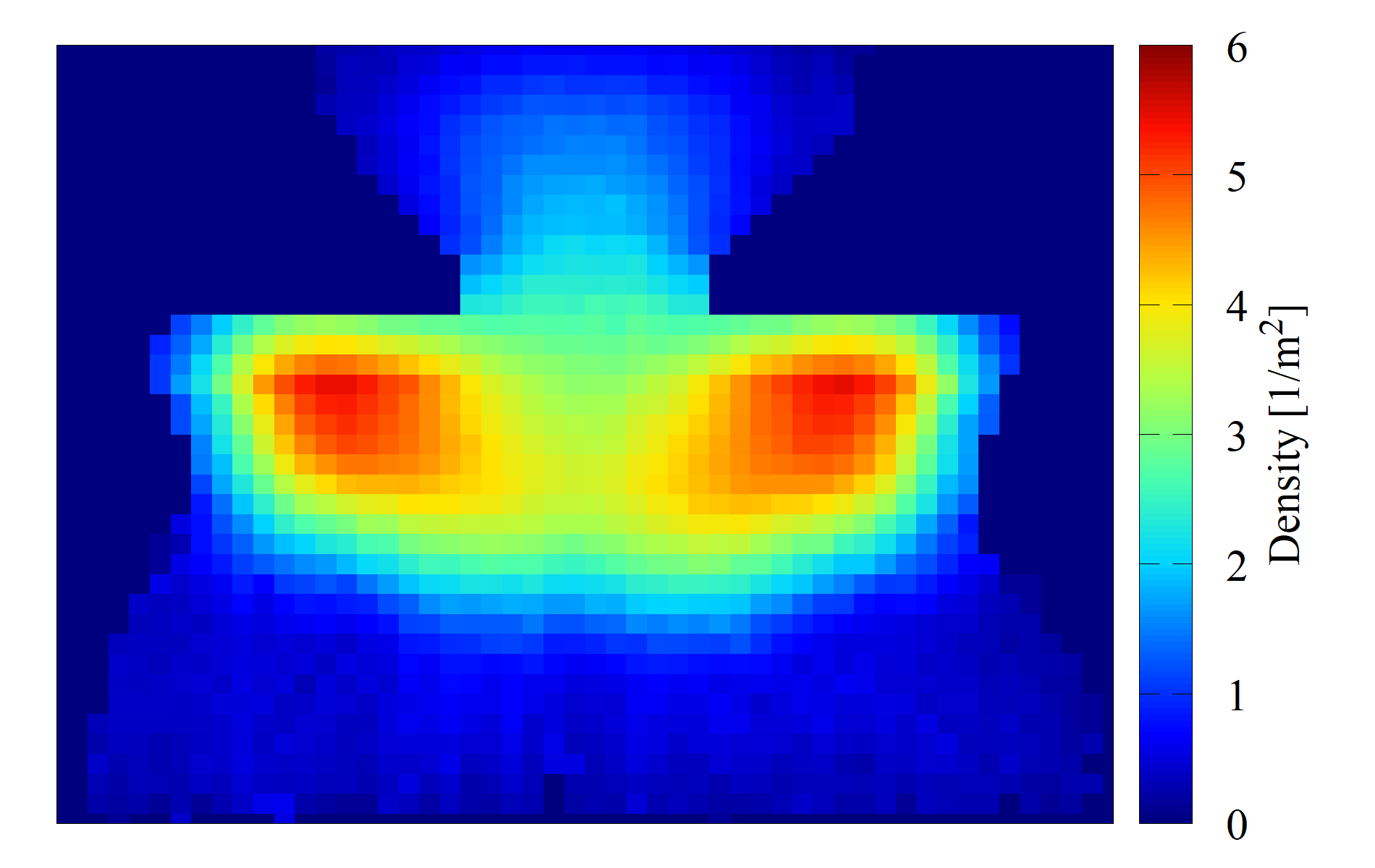}}
\subfigure[]{\includegraphics[width=0.45\textwidth]{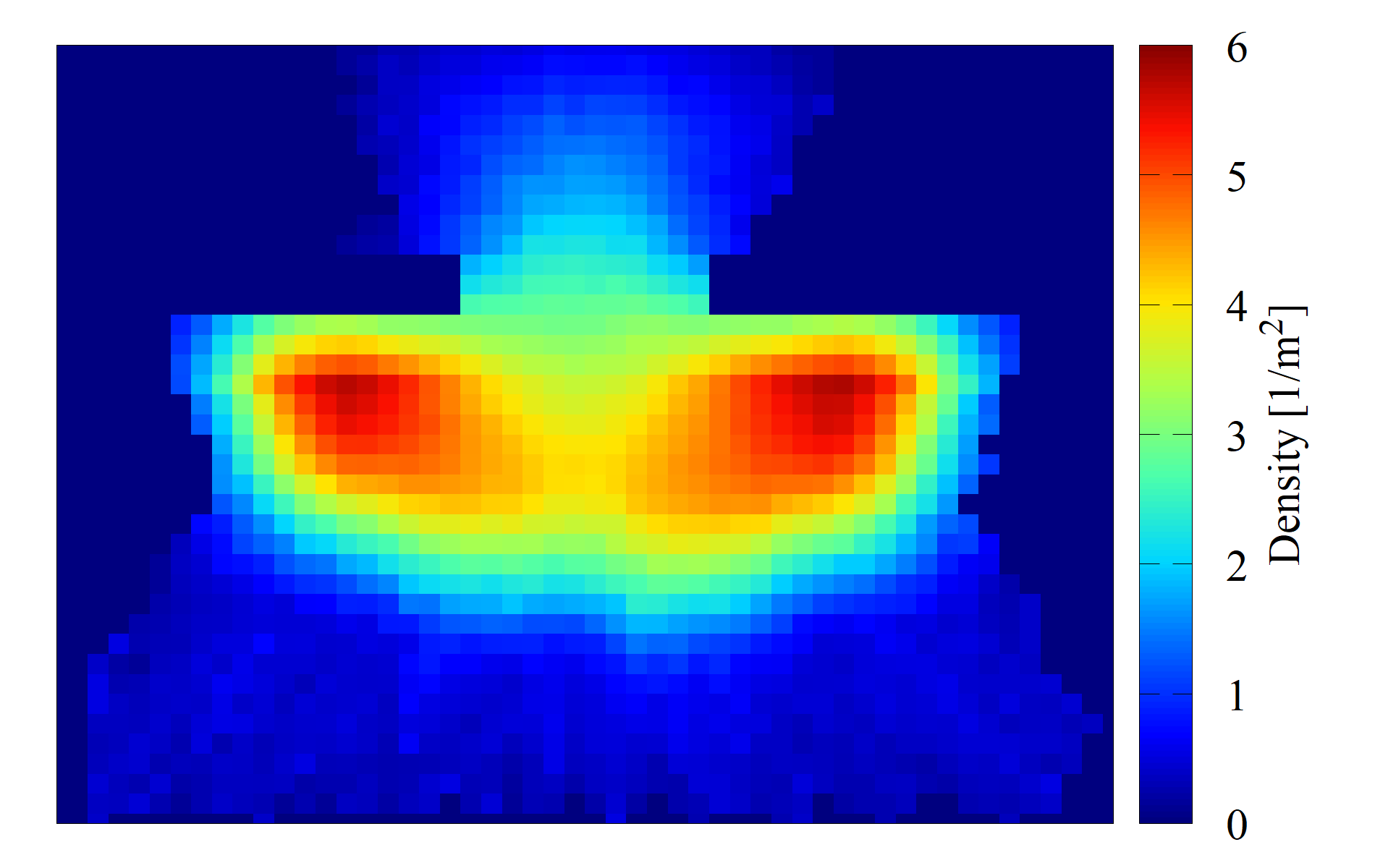}}
\caption{Cumulative mean density at a steady state for the scenario with only individuals (a) and the one with 50\% presence of dyads (b).}
\label{fig:bottleneck_CMD}
\end{figure}

Up to now, the model is calibrated to reproduce a plausible behaviour of dyads and it is validated for the simulation of uni-directional flow in a corridor. In this section the model will be used to study the effect of two-members groups in a periodical scenario representing a room with a bottleneck, as in Fig.~\ref{fig:bottleneck_scenario}. Again, we configure two simulation campaigns to study the phenomena: (a) a scenario with only individuals and (b) one with 50\% presence of dyads. 400 simulated pedestrians are generated in the scenario for each configuration of the bottleneck width and 5 iterations of 5000 steps are run to get a sufficient dataset. Again, output data are gathered only in a steady state of the system (time-step from 2000 to 5000). Fig.~\ref{fig:bottleneck_CMD} exemplifies two density maps for one iteration with bottleneck width of 4.0 m and highlights a slightly higher congestion in the scenario with dyads. This provides an explanation of the results describing the outflow of pedestrians from the bottleneck, which is calculated as $J = \frac{N}{t}$. The corresponding results are shown in Fig.~\ref{fig:bottleneck_results}. While the baseline scenario generates a specific flow at the bottleneck of about 2.1 ped/m*s, the scenario with dyads provides a sensibly lower result following the trend of about 1.8 ped/m*s observed in the dataset from Rupprecht et Al.~\cite{Rupprecht2011}.

\section{Conclusions}
The influence of cohesion among members of dyads and its aggregated effect in scenarios characterized by uni-directional flow has been investigated by applying a discrete model to two benchmark scenarios: a one way corridor and the egress from a room with a central bottleneck. The simulation model represents the dynamics of simulated group members with an adaptive mechanism, balancing the probability of movement according to the dispersion of the group; the cohesion mechanism is calibrated through two parameters $\kappa_c$ and $\delta$, acting on the utility of a movement, and according to data coming from a controlled experiments involving dyads. 

The investigated scenarios are simulated with two procedures, configuring a population composed of only individual pedestrians and one half composed of dyad members. In accordance with recent studies in the literature, simulated results show that the dyads negatively affects the dynamics, leading generally to a lower velocity and a lower pedestrian flow. The difference in the fundamental diagram observed in the scenario of the corridor is sensible and for the range of densities lower than 1.5 ped/m$^2$ is similar with the empirical data from the experiment used for the calibration (see~\cite{crocianiTGFcouplesExp} in this book). Results coming from the simulation of the bottleneck show again a negative effect of the presence of dyads, which in this case is even more apparent. The achieved trend and the difference between the two simulation scenarios are in line with other studies from the literature and, overall, suggests further research to investigate the microscopic behaviour of pedestrian groups and its aggregated effects on pedestrian dynamics.

\begin{acknowledgement}
This research has been supported by the Key Research and Development Program (2016YFC0802508), the Program of Shanghai Science and Technology Committee (16DZ1200106), the Specialized Research Fund for the Doctoral Program of Higher Education of China (20133402110009) the China Scholarship Council (CSC) and Fundamental Research Funds for the Central Universities (WK2320000035).
\end{acknowledgement}

%The reference style for working with bibtex is 

%alternatively, you can input the references manually. Keep to the format described in 
%Mathematical and Physical Sciences.pdf
%\input{referenc}
\end{document}